\documentclass[11pt]{article}
\usepackage[margin=1in]{geometry}
\usepackage{graphicx}
\usepackage{calrsfs}
\usepackage{cite}
\usepackage{amsfonts}
\usepackage{amsmath}
\DeclareMathAlphabet{\mathcal}{OMS}{cmsy}{m}{n}
\DeclareMathAlphabet{\pazocal}{OMS}{zplm}{m}{n}
\usepackage{xcolor}
\usepackage{lineno}
\usepackage{cleveref}
\usepackage{bm}

\def\fs{\rm 4e,4\mu,2e2\mu}
\def\mfourl{m_{4\ell}}
\def\mfourlp{m'_{4\ell}}
\def\dmfourl{\delta m_{4\ell}}
\def\mll{m_{\ell\ell}}
\def\pt{p_{\mathrm{T}}}
\def\ptdown#1{p_{ \mathrm{T}, #1}}
\def\ptup#1{p_{ \mathrm{T}}^{#1}}
\def\hatptup#1{\hat{p}_{ \mathrm{T}}^{#1}}
\def\dpt#1{\delta p_{\mathrm{T}, #1}}
\def\dptup#1{\delta p_{\rm T}^{\rm #1}}
\def\Dmass{{\mathcal D}_{\rm mass}}
\def\Dmassp{{\mathcal D}'_{\rm mass}}
\def\Dkinbkg{{\mathcal D}^{\rm kin}_{\rm bkg}}
\def\zplusx{\mathrm{Z}+\mathrm{X}}
\def\z#1{\mathrm{Z}_{#1}}
\def\decayobs{\vec{\Omega}^{\mathrm{H} \rightarrow 4\ell}}
\def\Pqqbkg{P^{\mathrm{q\bar{q}}}_{\mathrm{bkg}}}
\def\Pggsig{P^{\mathrm{gg}}_{\mathrm{sig}}}

\def\hzzfourl{\mathrm{H} \rightarrow \mathrm{Z}\mathrm{Z} \rightarrow 4\ell}
\def\haa{\mathrm{H} \rightarrow \gamma \gamma}
\def\ztoll{{\rm Z} \rightarrow \ell^{-}\ell^{+}}
\def\qqzz{\mathrm{q \bar{q} \rightarrow ZZ}, \mathrm{Z}\gamma^{*}}
\def\ggzz{\mathrm{g} \mathrm{g} \rightarrow \mathrm{Z}\mathrm{Z}, \mathrm{Z}\gamma^{*}}
\def\qqgg{\mathrm{q} \bar{\mathrm{q}} / \mathrm{g} \mathrm{g} \rightarrow \mathrm{Z}\mathrm{Z}, \mathrm{Z}\gamma^{*}}

\def\mhvalrun1{125.09 \pm 0.24\ \mathrm{GeV}}
\def\mhvalrunonelong{125.09 \pm 0.24\ (\pm 0.21[\mathrm{stat.}] \pm 0.11[\mathrm{sys.}])\ \mathrm{GeV}}

\def\mhvalshort{125.26 \pm 0.21\ \mathrm{GeV}}
\def\mhvallong{125.26 \pm 0.21\ (\pm 0.20[\mathrm{stat.}] \pm 0.08[\mathrm{sys.}])\ \mathrm{GeV}}
\def\massof#1#2{m({\mathrm{#1}_{#2}})}
\def\fb#1{{#1}\ \mathrm{fb}^{-1}}

\newcommand{\Likeli}{\pazocal{L}}

\def\beq{\begin{equation}}
\def\eeq#1{\label{#1}\end{equation}}
\def\eeqn{\end{equation}}

\def\beqa{\begin{eqnarray}}
\def\eeqa#1{\label{#1}\end{eqnarray}}
\def\eeqan{\end{eqnarray}}

\let\bar=\overbar

\def\eg{{\it e.g.}}

\def\Dslash{\not{\hbox{\kern-4pt $D$}}}
\def\dslash{\not{\hbox{\kern-2pt $\del$}}}

\def\mz{m_{\mathrm{Z}}}

\def\mh{m_\mathrm{H}}

\def\msb{{\bar{\ssstyle M \kern -1pt S}}}

\def\Title#1{\begin{center} {\Large {\bf #1} } \end{center}}
\def\Author#1{\begin{center} {\normalsize {\sc #1} } \end{center}}
\def\Institution#1{\begin{center} {\normalsize {\it #1} } \end{center}}
\def\Abstract#1{\noindent {\normalsize {\bf Abstract:} {\normalfont #1}}}
\def\Conference{\vspace{4mm}\begin{raggedright} {\normalsize {\it Talk presented at the 2019 Meeting of the Division of Particles and Fields of the American Physical Society (DPF2019), July 29--August 2, 2019, Northeastern University, Boston, C1907293.} } \end{raggedright}\vspace{4mm}}

\begin{document}
%
%

\Title{Higgs boson mass measurement\\ 
using \bm{$\hzzfourl$} decays at CMS}

\Author{Jake Rosenzweig on behalf of the CMS Collaboration}

\Institution{Department of Physics\\ University of Florida, Gainesville, FL, USA}

\Abstract{
A summary of the methods used to make a precision measurement of the Higgs boson mass is presented. 
The final mass value for the Higgs boson is measured to be 
$\mh = \mhvalshort$.  
This analysis considers the $\hzzfourl$ channel 
($\ell = e, \mu$), using proton-proton collision data collected in 2016 corresponding to an integrated luminosity of $\fb{35.9}$ at $\sqrt{s} = 13$ TeV
by the CMS experiment at the LHC.
A mass constraint is imposed on the invariant mass of the two leptons coming from the mostly on-shell Z boson to 
refit the lepton momenta and, hence, improve the measurement of the Higgs boson mass, per-event.
The mass of the Higgs boson is extracted using a three-dimensional likelihood fit, which uses three observables per-event:
(1) the refitted four-lepton invariant mass ($\mfourlp$), 
(2) the refitted four-lepton mass uncertainty ($\Dmassp$), and 
(3) a matrix element-based kinematic discriminant ($\Dkinbkg$). }

\Conference 
%
%
\section{Introduction}

In 2015, two major particle physics collaborations, CMS and 
ATLAS, jointly published measurements on the mass of the recently-discovered Higgs boson~\cite{2012atlas,2012cms}.
These results used proton-proton collision data delivered by the LHC at a center-of-mass energy of 13 TeV during Run 1 (2011 and 2012) corresponding to a total integrated luminosity of approximately $\fb{25}$ per collaboration.
The Higgs boson mass $(\mh)$ measurement during Run 1 using the combined results from both collaborations was $\mh = \mhvalrun1$.
Two Higgs boson decay channels were analyzed to obtain the final mass result: 
(1) $\haa$ and (2) $\hzzfourl$ 
(where $\ell = e, \mu$).
The individual mass measurements for each decay channel, including statistical and systematic uncertainties, and their combined results per collaboration, are given in Table~\ref{tab:table_higgsmassmeas}~\cite{PRL_114}.
\begin{table}[phtb]
\centering
    \includegraphics[width=0.98\textwidth]{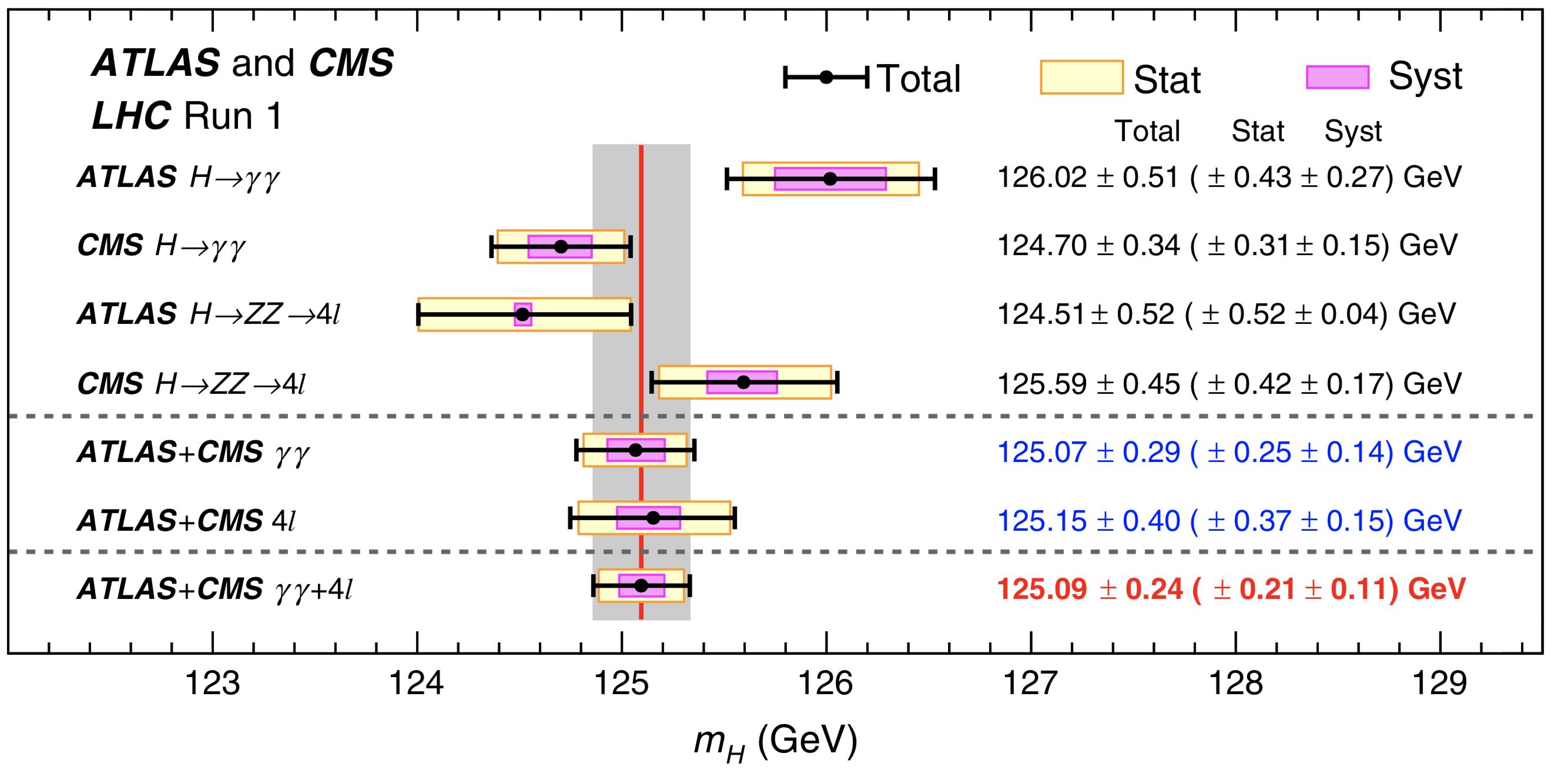}
    \caption{
        The Higgs boson mass measurement results from the ATLAS and CMS Collaborations during the LHC Run 1 using two different decay channels in combination.
        The statistical (yellow bands), systematic (pink bands), and total (black bars) uncertainties are shown.
        The central value (red line) and the corresponding total uncertainty (gray bar) for the final combined mass measurement are shown at the bottom in red text~\cite{PRL_114}.
    }
    \label{tab:table_higgsmassmeas}
\end{table}

In 2017, the precision on the aforementioned Run 1 mass measurement was superseded by CMS by considering only the $\hzzfourl$ channel and using 2016 Run 2 data, corresponding to a total integrated luminosity of $\fb{35.9}$.
The best value of $\mh$ is determined by minimizing a three-dimensional (3D) likelihood fit of the following three observables:
(1) the refitted four-lepton invariant mass ($\mfourlp$), 
(2) the refitted four-lepton mass uncertainty ($\Dmassp$), and 
(3) a matrix element-based kinematic discriminant ($\Dkinbkg$).
Each observable is evaluated on a per-event basis and uses refitted lepton transverse momenta ($\pt$) by performing a mass constraint on the more on-shell Z boson, which ultimately improves the precision of the 3D likelihood fit.

The Higgs boson mass resonance in the four-lepton state enables a precision measurement of $\mh$ due to a large signal-to-background ratio and because of its intrinsically narrow width.
The background processes consist of two kinds: 
(1) irreducible $\qqgg$ processes which skip Higgs boson production altogether, and
(2) reducible $\zplusx$, in which a Higgs boson decays into a Z boson and X, where X is typically a heavy flavor jet which decays into secondary leptons. These leptons are misidentified as prompt leptons and therefore this process is considered background.
This $\zplusx$ region is estimated using data-driven methods.
\section{Observables}
\label{sec:observables}

Three observables are used to extract the Higgs boson mass using a three-dimensional likelihood fit that scans over $\mh$. 
Each observable is defined in the following subsections. 
The observables are: 
\begin{enumerate}
\item the refitted four-lepton invariant mass $(\mfourlp)$, 
\item the refitted four-lepton mass uncertainty on an event-by-event basis $(\Dmassp)$,
\item a matrix element-based kinematic discriminant $(\Dkinbkg)$. 
\end{enumerate}
It should be noted that an observable with a prime symbol ($'$) indicates that it has been updated using the refitted lepton $\pt$ values as described in Section~\ref{sec:mass_constraint}. 

\subsection{First Observable: Four-Lepton Invariant Mass $\big(\mfourlp\big)$}
\label{subsec:mfourl}

\begin{figure}[phbt]
\centering
\includegraphics[width=0.98\textwidth]{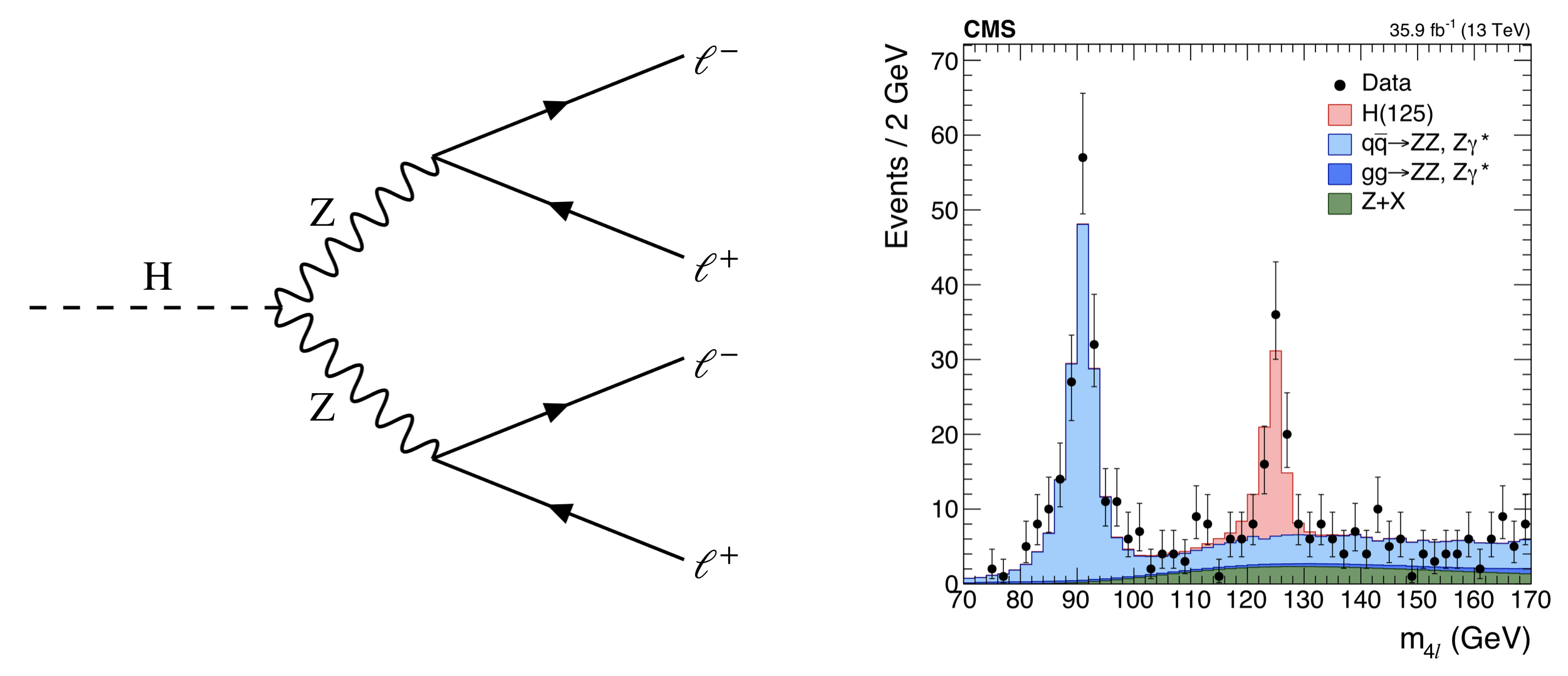}
    \caption{
        (Left) The Feynman diagram representing the $\hzzfourl$ process.
        (Right) The four-lepton invariant mass distribution. The Higgs boson mass resonance (salmon color) stands well above the $\qqzz$ (light-blue), $\ggzz$ (dark-blue), and $\zplusx$ (green) background processes~\cite{HIG-16-041}.}
    \label{fig:feyn}
\end{figure}

The topology of the $\hzzfourl$ process is shown in Figure~\ref{fig:feyn} (Left). 
A Higgs boson decays into two Z bosons, one of which is mostly on-shell $(\z1)$, while the other Z boson is mostly off-shell, $\z2$.
Each Z boson decays to opposite-sign, same-flavor leptons, 
$\ztoll$ ($\ell = e,\mu$), which yields the following final states: 4e, 4$\mu$, and 2e2$\mu$.
For each final state category, the four-lepton invariant mass ($\mfourl$), is evaluated per-event. 
The distribution of $\mfourl$ is shown in Fig.~\ref{fig:feyn} (Right)~\cite{HIG-16-041}.
The $\mfourl$ observable gets updated to $\mfourlp$ by a kinematic refitting of the lepton $\pt$ values using a $\z1$ mass constraint as described in Section~\ref{sec:mass_constraint}.

\subsection{Second Observable: Relative Mass Uncertainty $\big(\Dmassp\big)$}
\label{subsec:relmassunc}

Per-event, the evaluation of $\mfourl$ comes with an associated uncertainty ($\dmfourl)$ which is a function of the transverse momenta $(\ptdown{\ell_k})$ and the corresponding uncertainties $(\dpt{\ell_k})$ of the four outgoing leptons:
\begin{linenomath*}
\begin{equation}
    \dmfourl = 
    F\Big(
    \{p_{T,\ell_1},\delta p_{T,\ell_1}\}
    ,\cdots, 
    \{p_{T,\ell_4},\delta p_{T,\ell_4}\}\Big).
\label{eq:dmfourl}
\end{equation}
\end{linenomath*}
This value is obtained via an error propagation technique which smears
$\ptdown{\ell_k}$ by its corresponding $\dpt{\ell_k}$, lepton-by-lepton:
\begin{linenomath*}
\begin{equation}
   \ptdown{k} \rightarrow \ptdown{k} + \dpt{k}
   = p'_{\mathrm{T},\ell_k}
   \label{eq:smear}
\end{equation}
\end{linenomath*}
where $p'_{T,\ell_k}$ is the smeared transverse momentum of lepton $k$. 
An increased precision on the measurement of $\ptdown{\ell_k}$ (decreased $\dpt{\ell_k}$)
leads to an increased precision on $\mfourl$ (decreased $\dmfourl$).
The $\dpt{\ell_k}$ values are corrected to minimize $\dmfourl$ using lepton $\pt$ error correction factors $(\lambda)$:
\begin{linenomath*}
\begin{equation}
    \delta p^{\mathrm{corr}}_{{\rm T},\ell_k} = \lambda \times \dpt{\ell_k}.
\label{eq:lambda}
\end{equation}
\end{linenomath*}
These $\lambda$ values are evaluated by simulating $\ztoll$ ($\ell = e,\mu$) events and by extracting the resolution ($\sigma$) of the Z mass resonance from the $\mll$ distribution. 
This mass distribution is fit using a Breit-Wigner probability density function (pdf) convoluted with a Crystal Ball (CB) pdf and a decaying exponential pdf.
The fit then extracts the relevant parameters for each pdf, holds all of them fixed, except the resolution of the CB pdf ($\sigma_{\rm CB}$) which gets replaced by:
\begin{linenomath*}
\begin{equation*}
    \sigma_{\rm CB} \rightarrow \lambda \times \delta m_{\rm \ell\ell}
\end{equation*}
\end{linenomath*}
where $\lambda$ is the sought-after correction factor to be used in Eq.~\ref{eq:lambda}.
A second fit, which is conditional on the per-event uncertainty of the dilepton invariant mass $(\delta \mll)$, finds the single, optimal $\lambda$ value for this simulated sample of $\ztoll$ events.

\begin{figure}[pbth]
\centering
\includegraphics[height=3.0in]{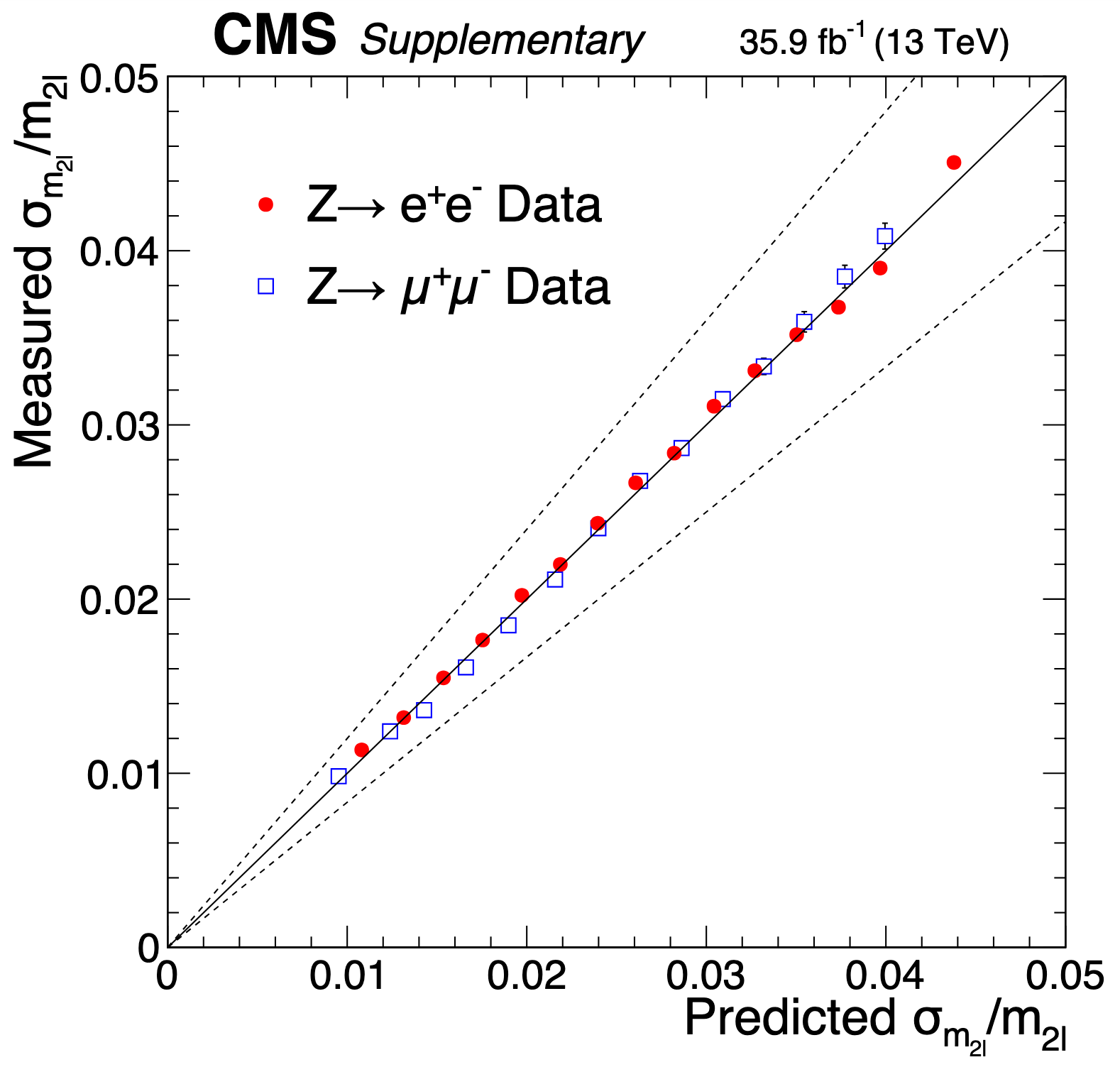}
    \caption{
    A closure test is used to validate the lepton $\pt$ error correction values $(\lambda)$. The measured relative mass uncertainty values (Measured $\sigma_{\rm m_{2\ell}} / m_{\rm 2\ell})$ for simulated $\ztoll$ events are compared with the predicted relative mass uncertainty values. 
    Events are binned according to their predicted relative mass uncertainty such that approximately an equal number of events is found in each bin.
    Validation of the $\lambda$ values is confirmed in each bin, for both final states (red dots for $2e$, open squares for $2\mu$), since the measured values are tightly bound around the 1-to-1 line (solid black). The dashed lines represent a 20\% systematic uncertainty on the resolution during Run 1~\cite{HIG-16-041}.
    } 
    \label{fig:mz_fit}
\end{figure}

It is important to note that, since different parts of the CMS detector have different lepton $\pt$ resolutions, it is necessary to categorize each dilepton events into a specific kinematic bin, depending on the pseudorapidity $(\eta)$ and relative $\pt$ uncertainties $(\delta \pt/\pt)$ of the leptons detected by the various sub-detector regions~\cite{Chatrchyan_2008aa}.
Therefore, a $\lambda$ correction factor is evaluated for each kinematic bin and subsequently used to correct the lepton transverse momentum uncertainty in the corresponding region, per Eq.~\ref{eq:lambda}.
A closure test is performed to verify that the $\lambda$ values are performing as intended. 
As shown in Fig.~\ref{fig:mz_fit}~\cite{HIG-16-041}, the measured relative mass uncertainties of the dilepton events within a specific kinematic bin agree with the predicted values.
The $\delta p^{\mathrm{corr}}_{{\rm T},\ell_k}$ values are obtained per-event using Eq.~\ref{eq:lambda}. Next,
each ${p'_{T,\ell_k}}$ is individually updated using Eq.~\ref{eq:smear}.
Finally, the total $\dmfourl$ is evaluated using the updated ${p'_{T,\ell_k}}$ values per-event using Eq.~\ref{eq:dmfourl}.

It is useful to normalize $\dmfourl$ with the corresponding $\mfourl$ value and define the per-event relative mass uncertainty ($\Dmass$) as:
\begin{linenomath*}
\begin{equation*}
    \Dmass = \frac{ \delta m_{4\ell} }{ m_{4\ell}
    }.
\end{equation*}
\end{linenomath*}
As will be described in Section~\ref{sec:mass_constraint}, the $\Dmass$ gets updated using refitted lepton $\pt$ values and is promoted to a refitted, per-event relative mass uncertainty ($\Dmassp$). 
\subsection{Third Observable: Kinematic Discriminant $\big(\Dkinbkg\big)$}
\label{subsec:kindisc}

\begin{figure}[phbt]
\centering
\includegraphics[width=0.49\textwidth]{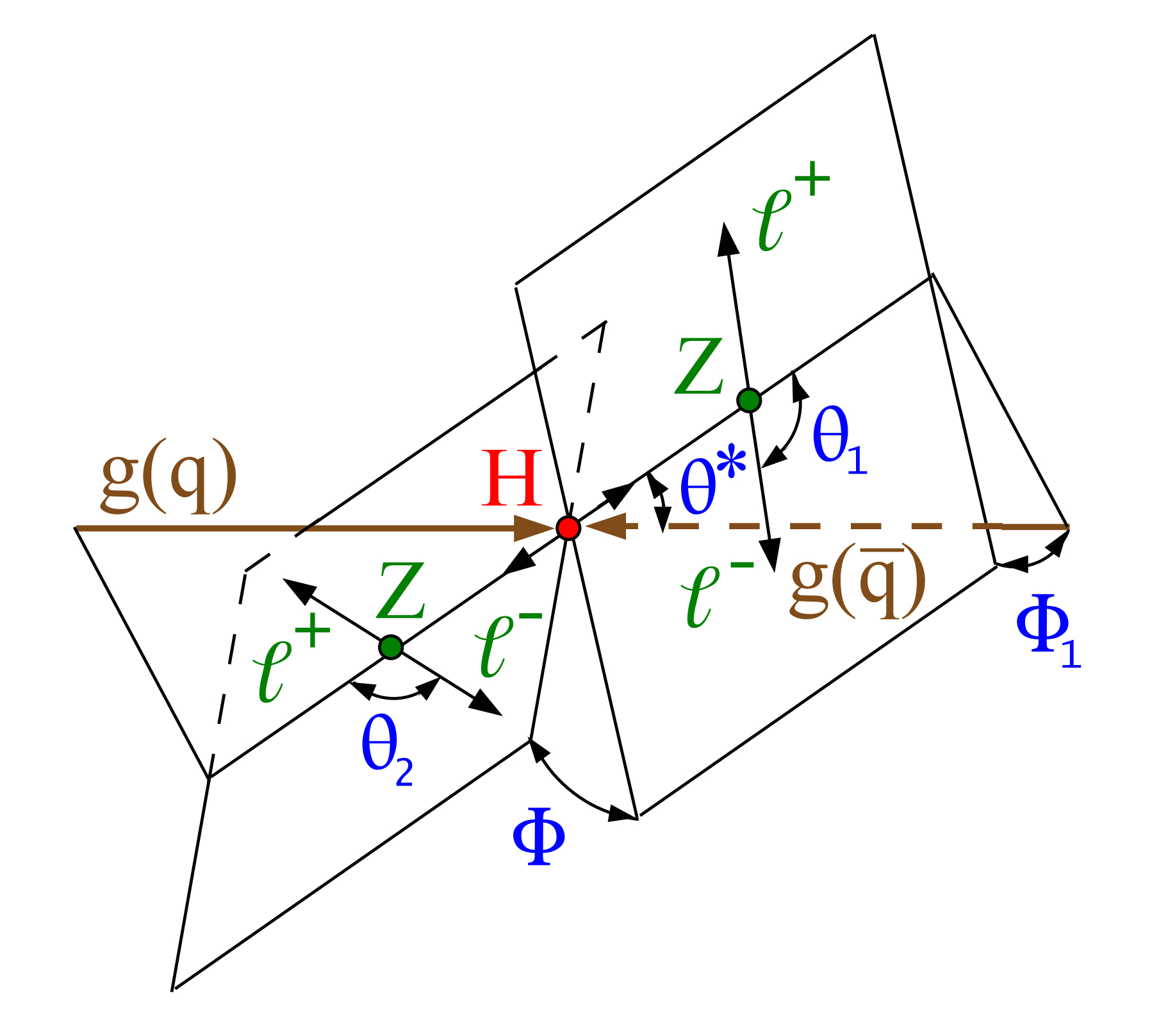}
\includegraphics[width=0.49\textwidth]{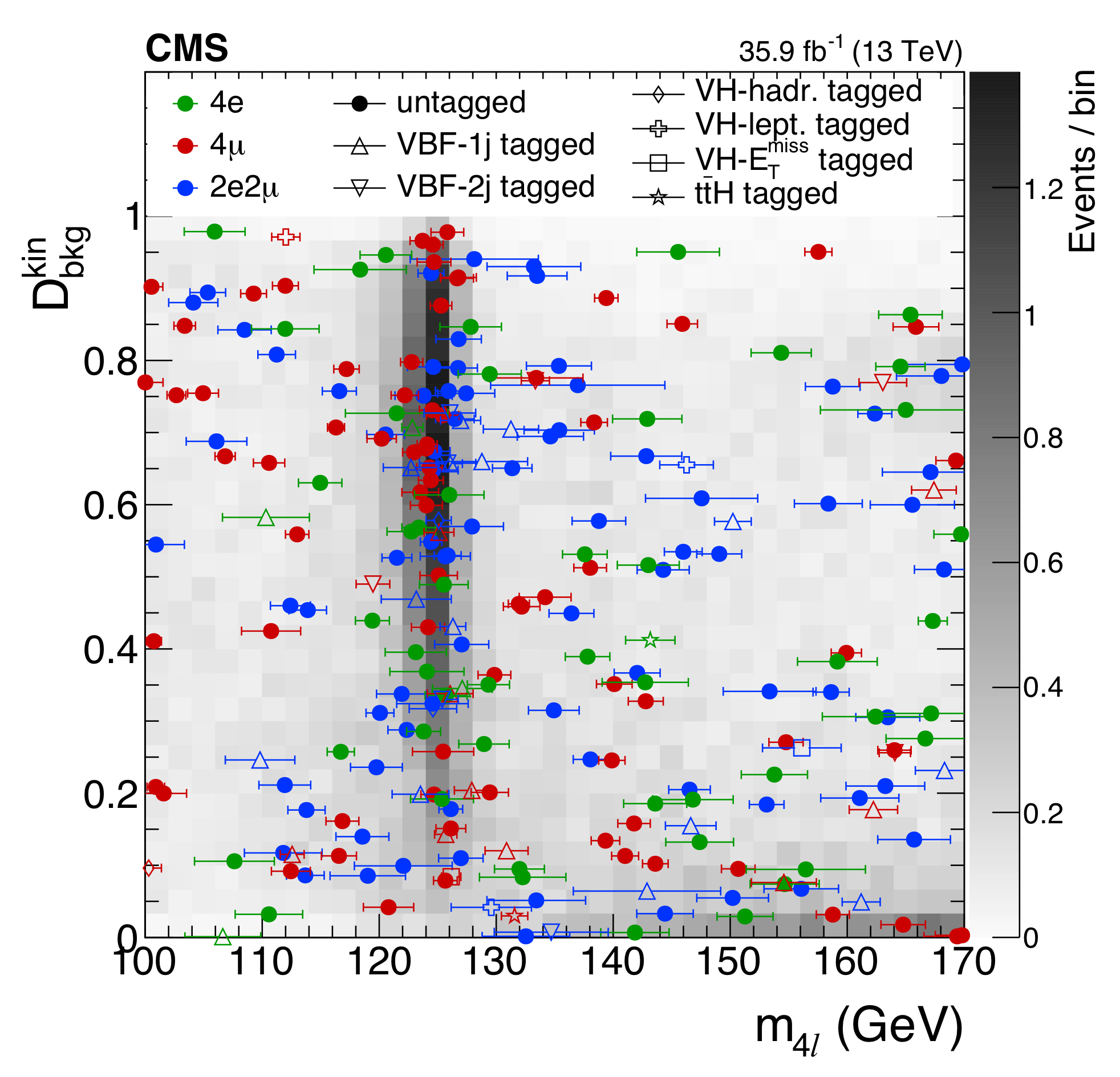}
\caption{
     (Left) A ${\rm gg(q \bar{q})} \rightarrow \hzzfourl$ event, which shows the angular kinematical variables that go into $\decayobs$ and are then used to evaluate $\Dkinbkg$, per-event~\cite{Choudhury}.
    (Right) The $\Dkinbkg$ vs. $\mfourl$ is plotted in the range $100< \mfourl <170$ GeV, using $\Dmass$ as the relative mass uncertainty on each data point. The expected number of Higgs boson decay events (dark gray region near 125 GeV) agrees relatively well with the observed data (green, red, and blue points for each type of final state)~\cite{HIG-16-041}.
    }
\label{fig:Dkinbkg_plots}
\end{figure}

Each Higgs boson decay event, whether signal or background, carries certain kinematical information about the particles involved, \eg, the spins, momenta, and decay angles, of each particle.
Fig.~\ref{fig:Dkinbkg_plots} (Left) shows an example of such a Higgs boson decay and the associated kinematical variables involved~\cite{HIG-16-041}.
All the decay observables are then incorporated into a single variable $(\decayobs)$~\cite{gao,bolognesi,anderson}.
Then, a kinematic discriminant $(\Dkinbkg)$ can be constructed from $\decayobs$\ to better discriminate between signal and background events. 
The $\Dkinbkg$ is defined as:
\begin{linenomath*}
\begin{equation} 
\Dkinbkg = \Bigg[ 1 + \frac{
    \Pqqbkg (\decayobs|m_{4\ell}) }{
    \Pggsig (\decayobs|m_{4\ell})
}
\Bigg]^{-1}
\end{equation}
\end{linenomath*}
where $\Pqqbkg$ represents the probability density that an event is consistent with a 
$\qqzz$
background event, while
$\Pggsig$ represents the probability density that an event is consistent with a signal event.
Fig.~\ref{fig:Dkinbkg_plots} (Right)\cite{HIG-16-041} shows the three observables per-event, for simulation and data. 

\section{Kinematic Refit Using a $\z1$ Mass Constraint}
\label{sec:mass_constraint}

\begin{figure}[ptb]
\centering
    {\includegraphics[width=0.49\textwidth]
    {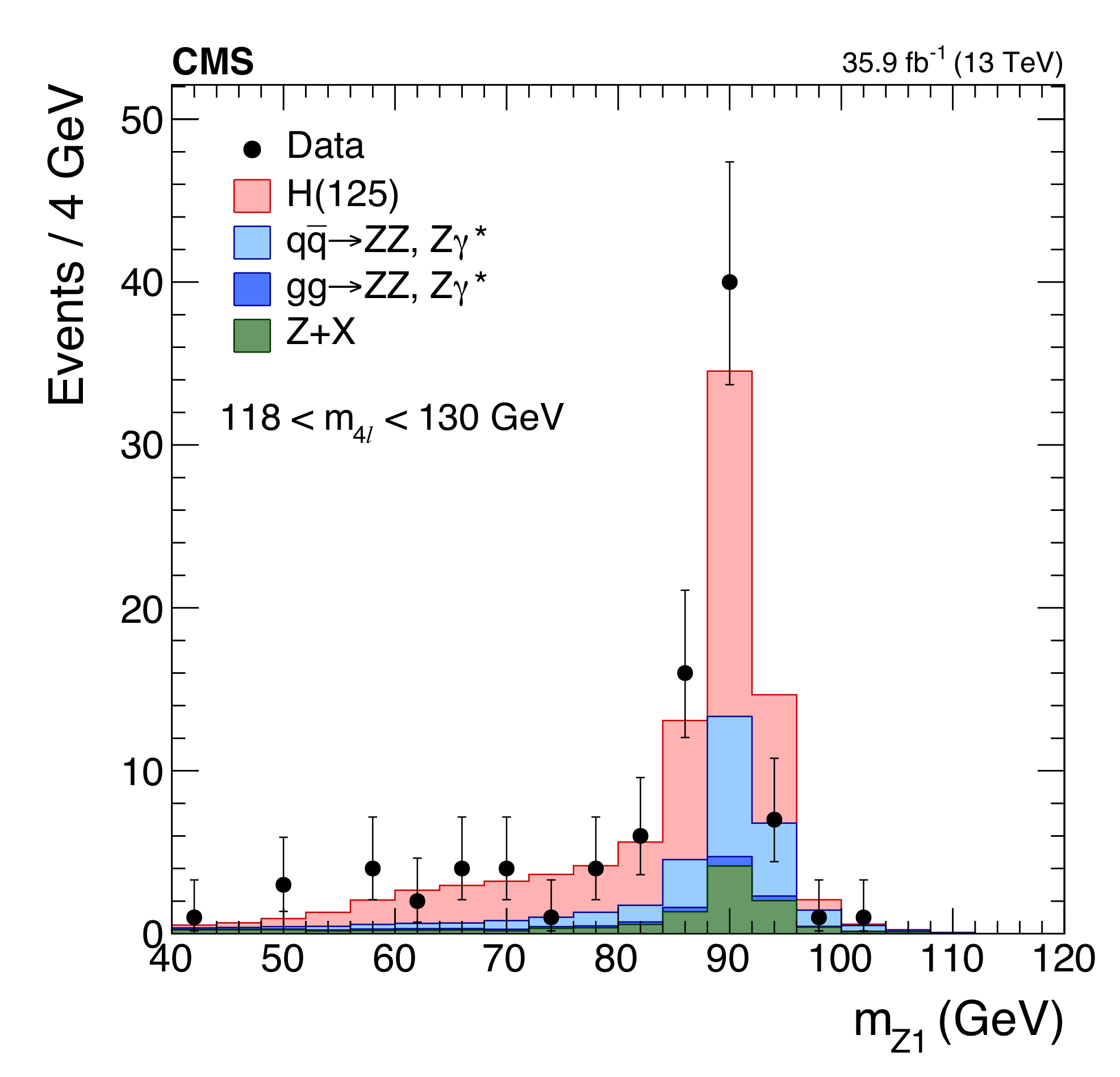}}
    {\includegraphics[width=0.49\textwidth]
    {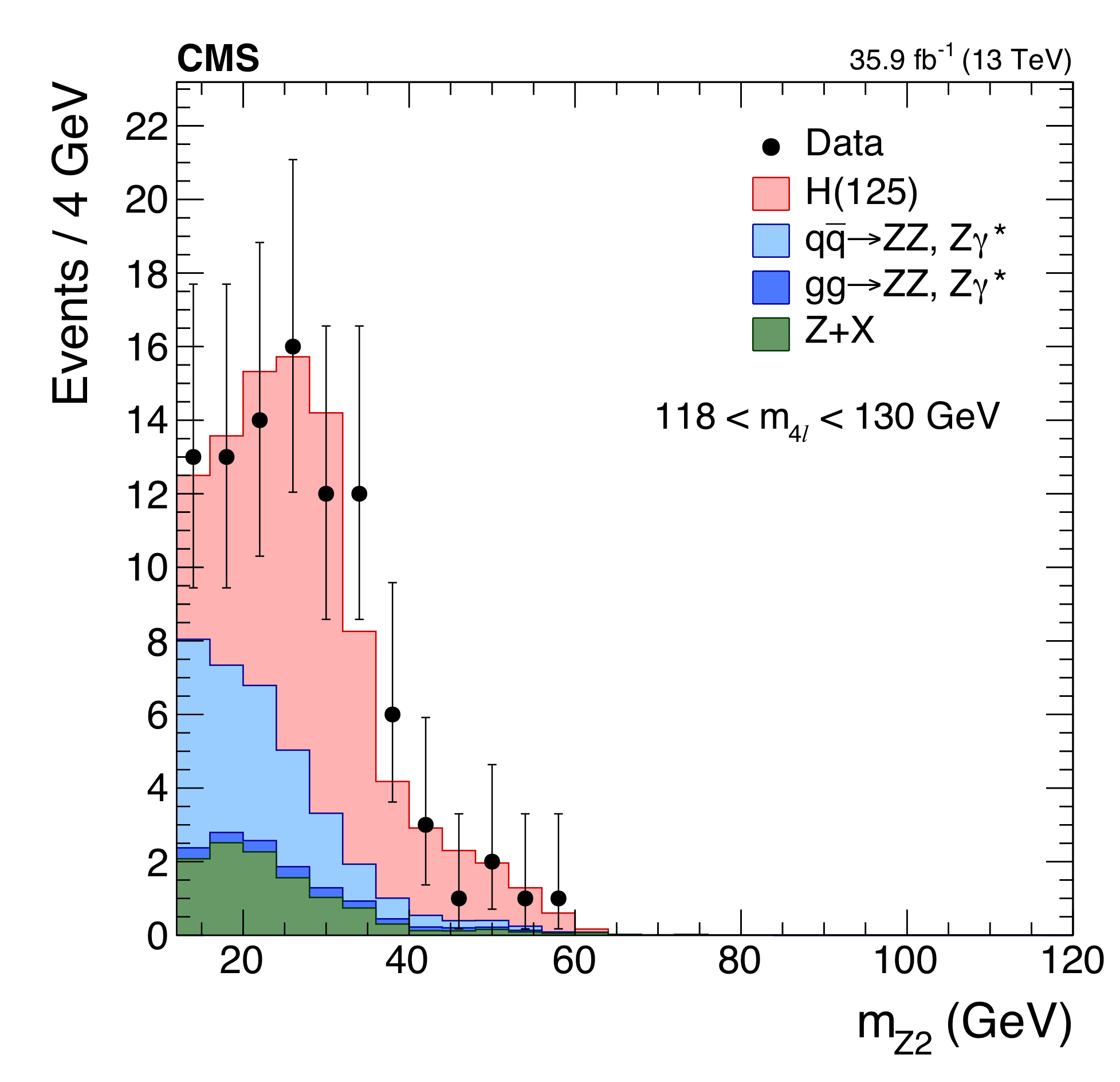}}
\caption{
(Left) Invariant mass distribution of $\ztoll$ events whose dilepton invariant mass is closer to the PDG value of the Z boson (91.188 GeV) than the other dilepton pair invariant mass. 
The $\z1$ has a narrow width and is mostly on-shell.
(Right) Invariant mass distribution of the dilepton events whose mass is farther from the Z boson PDG value as compared to the other dilepton pair.
In both mass distributions, it is required that $118 < \mfourl < 130$ GeV~\cite{HIG-16-041}.}
\label{fig:z1z2}
\end{figure}

The Higgs boson mass measurement can achieve even better precision by using knowledge of the mass and line shape of the mass resonance of the more on-shell Z boson $(\z1)$.
As shown in Fig.~\ref{fig:z1z2} (Left) the $\z1$ is mostly on-shell and therefore allows for precise fitting of its line shape, whereas the $\z2$ is mostly off-shell~\cite{HIG-16-041}.
Because of the well-defined line shape of the $\z1$ mass resonance, and the fact that the narrow width of the resonance is similar to that of the detector resolution, a mass constraint is applied which allows for a refitting of the dilepton $\pt$ values originating from the $\z1$, on a per-event basis.
The mass constraint is used in maximizing the likelihood of the refitted lepton $\pt$ values:
\begin{linenomath*}
\begin{equation}
    \Likeli( \hatptup{1},\hatptup{2} |  
    \ptup{1}, \ptup{2}, \dptup{1}, \dptup{2}) 
    = 
    {\rm Gauss}( \ptup{1} | \hatptup{1}, \dptup{1} )
    {\rm Gauss}( \ptup{2} | \hatptup{2}, \dptup{2} )
    \Likeli (m_{12} | \mz, \mh)
  \label{eq:mass_constraint}
\end{equation}
\end{linenomath*}
where $\ptup{1}$ and $\ptup{2}$ are the reconstructed $\pt$ values of the two leptons originating from the $\z1$, 
$\delta {\ptup{1}}$ and $\delta {\ptup{2}}$ are the corresponding per-event lepton $\pt$ uncertainties,
$\hatptup{1}$ and $\hatptup{2}$ are the refitted $\pt$ values of the two leptons, 
and $m_{12}$ is the invariant mass calculated using the refitted $\pt$ values. 
Here, $\mz$ is set to the PDG value for the Z boson mass (91.188 GeV), 
$\mh$ is set to 125 GeV corresponding to the mass use in simulated $\hzzfourl$ events,
and lastly the $\Likeli (m_{12} | \mz, \mh)$ is the mass constraint term on the $\massof{Z}{1}$. 

\begin{figure}[phbt]
\centering
    {\includegraphics[width=0.32\textwidth]{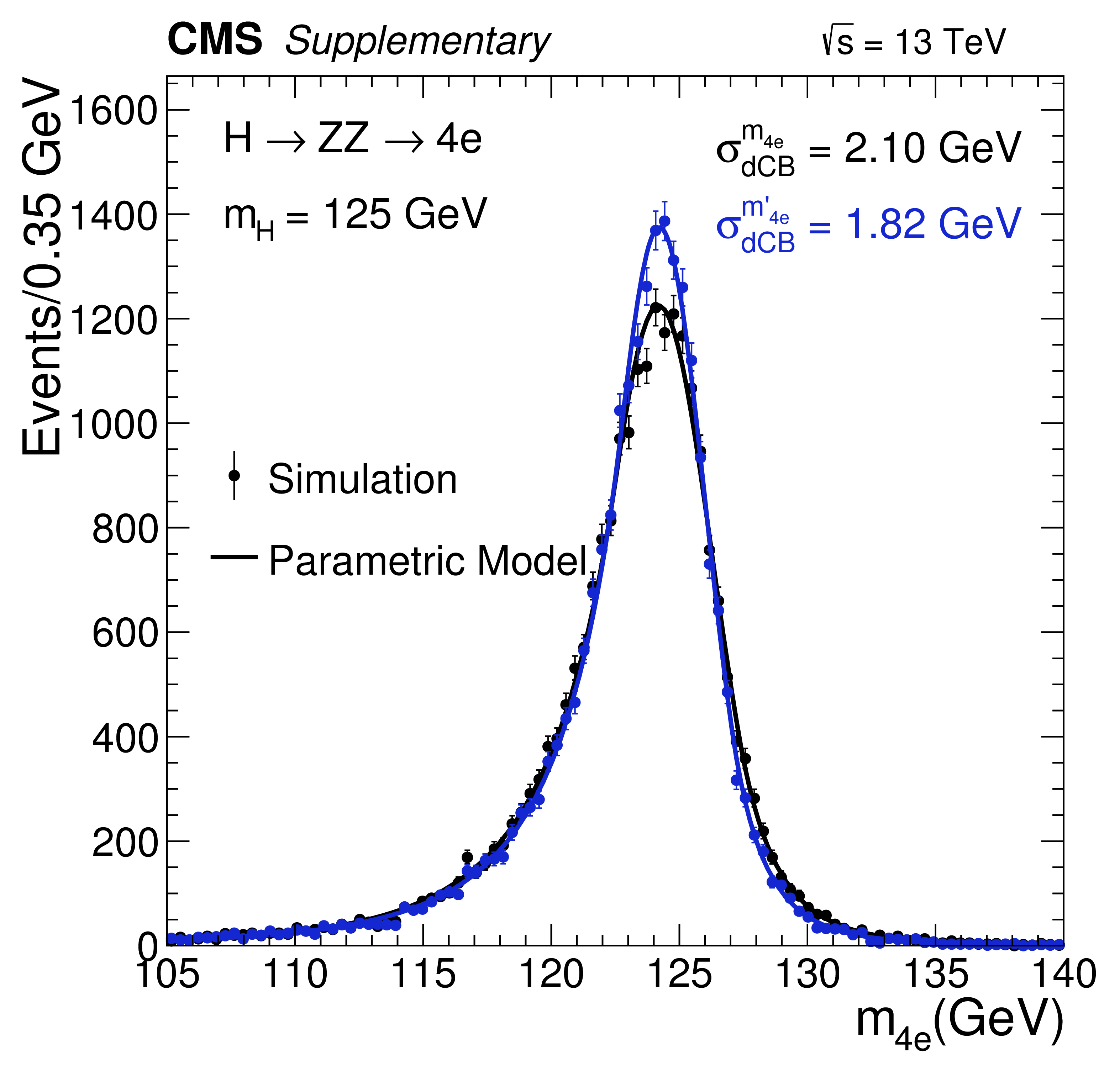}}
    {\includegraphics[width=0.32\textwidth]
    {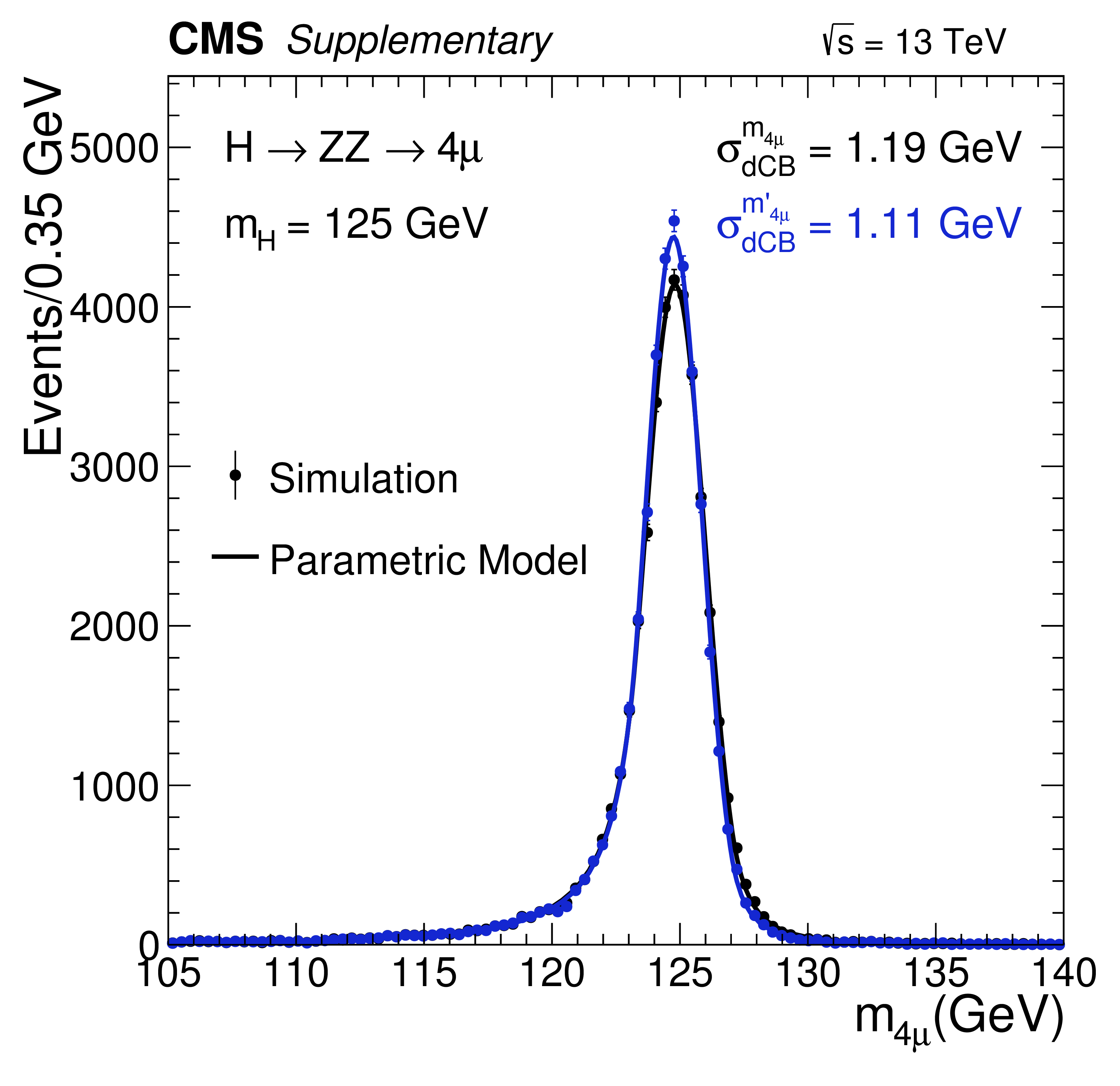}}
    {\includegraphics[width=0.32\textwidth]
    {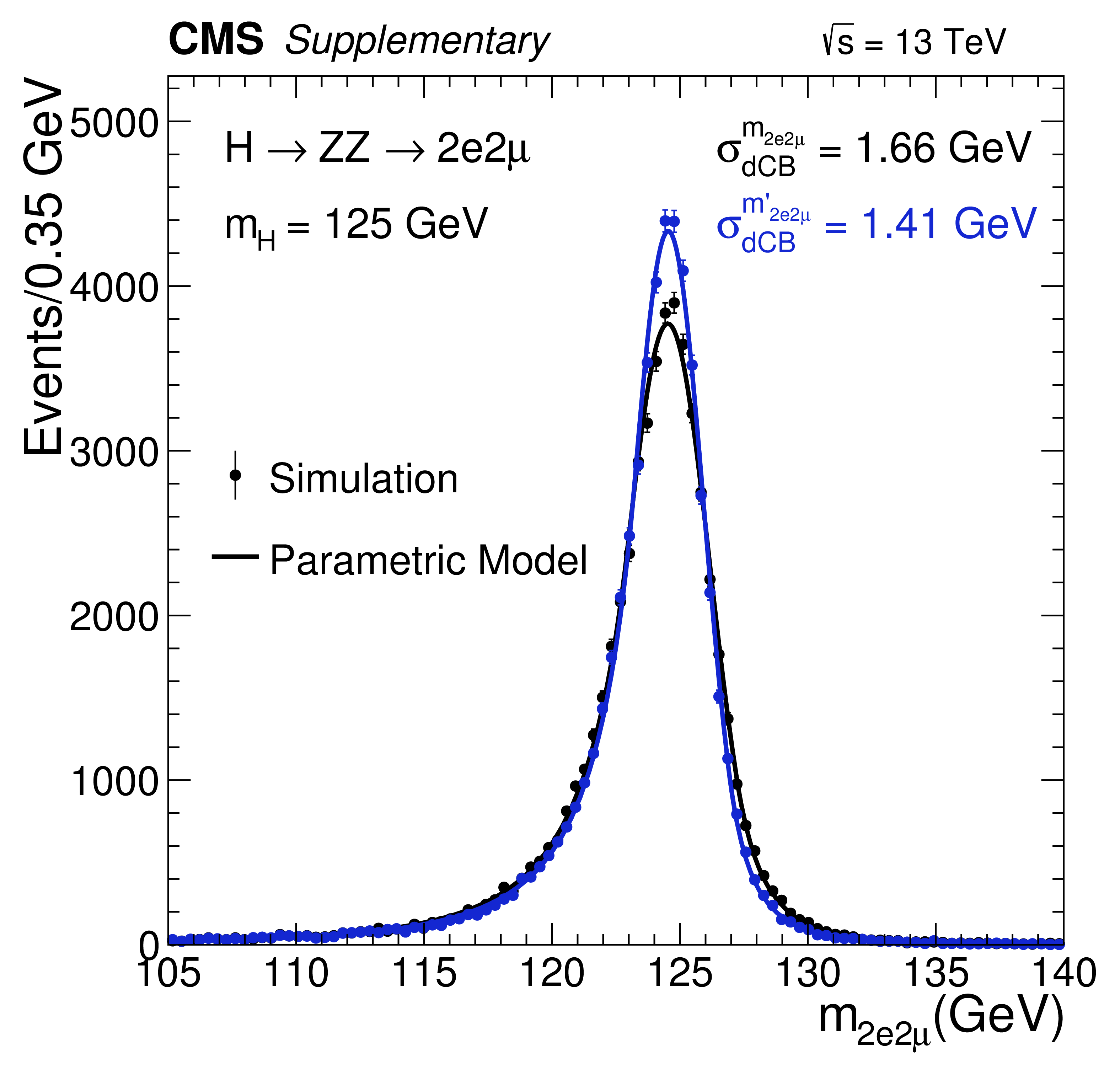}
    \caption{
    Distributions of the $\mfourl$ in $\hzzfourl$ events using reconstructed lepton $\pt$ values (black line) and refitted $\pt$ values (blue line), separated by each final state: 4e (Left), 4$\mu$ (Center), 2e2$\mu$ (Right).
    The refitted distributions are obtained by imposing the mass constraint on the $\massof{Z}{1}$.
    Each curve is fit with a double Crystal Ball pdf and the resolution of the peak ($\sigma_{\rm dCB}$) is extracted.
    The resolution improvement between reconstructed and refitted distributions is 
    7\% in the $4\mu$ channel, 
    13\% in the $2e2\mu$ channel,
    and 15\% in the 4e channel~\cite{HIG-16-041}. 
    }
    \label{fig:m4l_refit}}
\end{figure}

Per event, the likelihood given by Eq.~\ref{eq:mass_constraint} is maximized by the mass constraint term and the newly-obtained refitted lepton $\pt$ values are used to update the first two observables $(\mfourl, \Dmass)$, which are then denoted by a prime ($'$):
\begin{linenomath*}
\begin{align*}
    \mfourl &\rightarrow \mfourlp\\
    \Dmass &\rightarrow \Dmassp.
\end{align*}
\end{linenomath*}
Using the $\massof{Z}{1}$ mass constraint to obtain refitted lepton $\pt$ values in this fashion provides greater precision on the final Higgs boson mass measurement, as shown in Fig.~\ref{fig:m4l_refit}~\cite{HIG-16-041}.

\section{Conclusions}

\begin{figure}[phtb]
\centering
\includegraphics[width=0.49\textwidth]{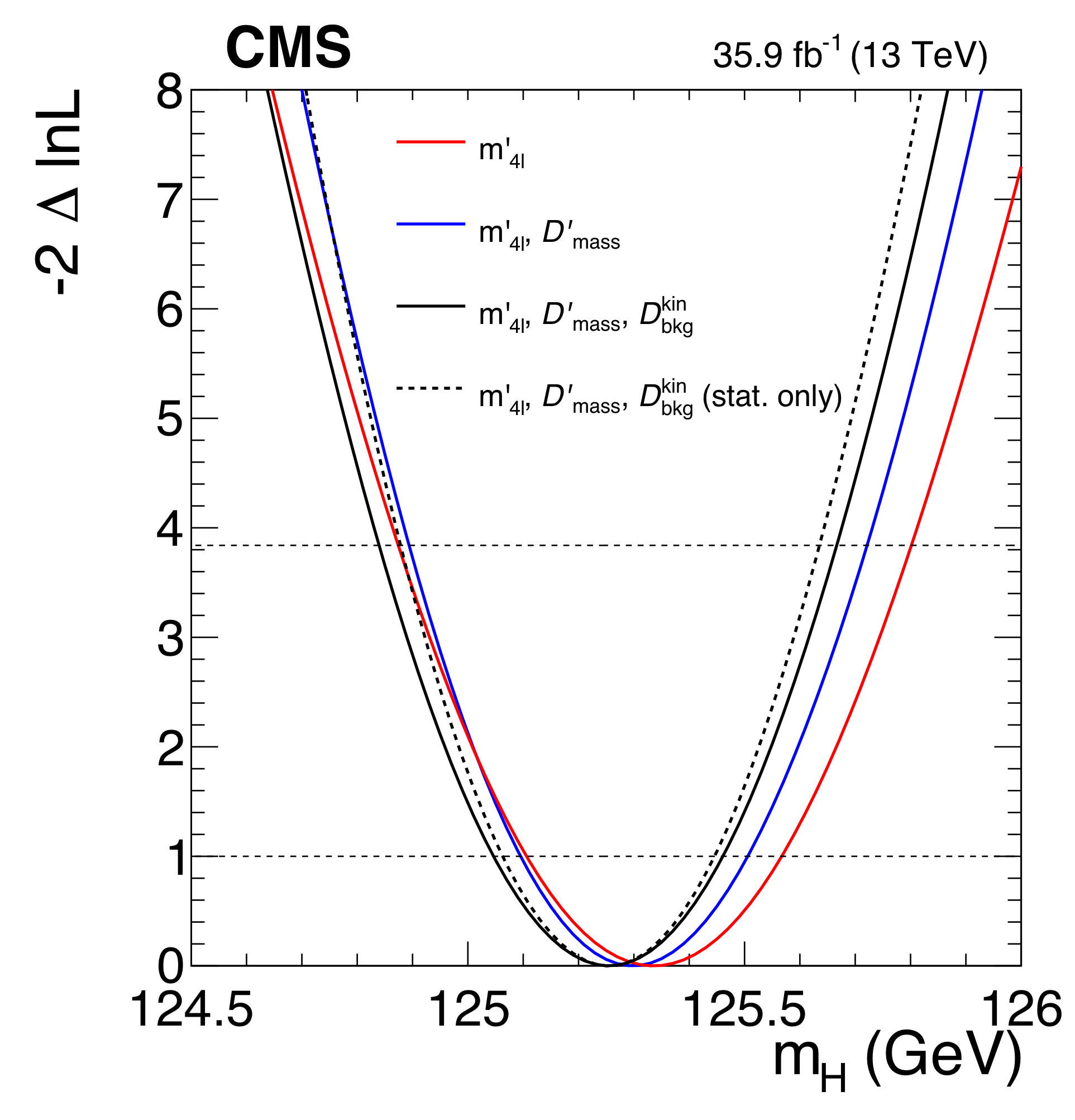}
\includegraphics[width=0.49\textwidth]{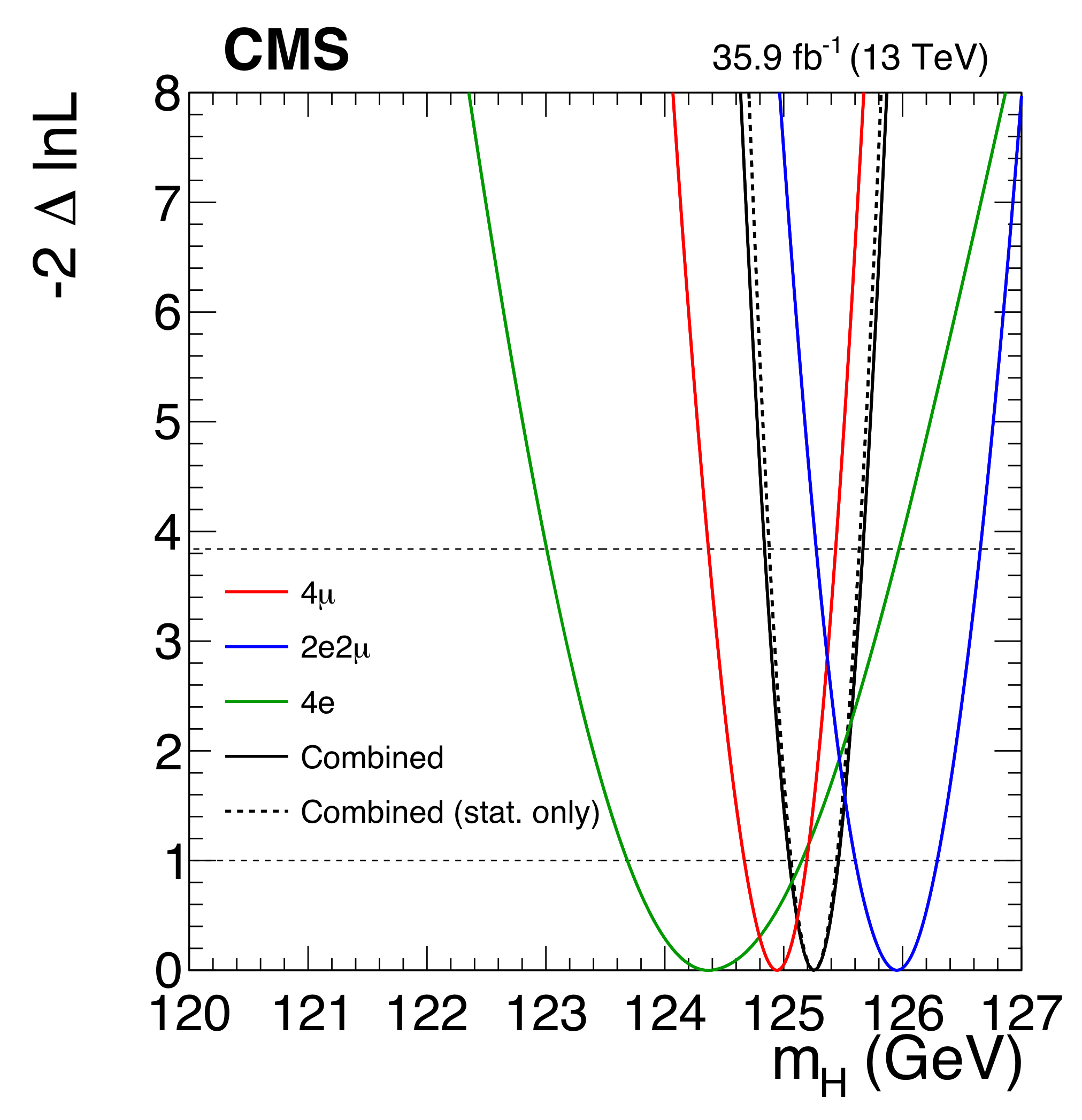}
\caption{
(Left) A likelihood scan of the Higgs boson mass using one (red), two (blue), or three (black) observables. 
The dashed black line includes only statistical uncertainty, whereas the solid black line includes both statistical and systematic uncertainty.
(Right) The 1D likelihood scan is split up into the different final states: $4\mu$ (green), $2e2\mu$ (blue), and 4e (green). Again, the dashed black line includes only statistical uncertainty, whereas the solid black line includes both statistical and systematic uncertainty
~\cite{HIG-16-041}.
}
\label{fig:likelihood_scan}
\end{figure}

The mass of the Higgs boson $(\mh)$ was extracted by a three-dimensional likelihood fit using three observables: $\mfourlp, \Dmassp, \Dkinbkg$. 
The data used in this measurement were collected by the CMS detector corresponding to an integrated luminosity of $\fb{35.9}$ from proton-proton collisions provided by the LHC during the 2016 Run.
\begin{table}[phtb]
\centering
\includegraphics[width=0.98\textwidth]{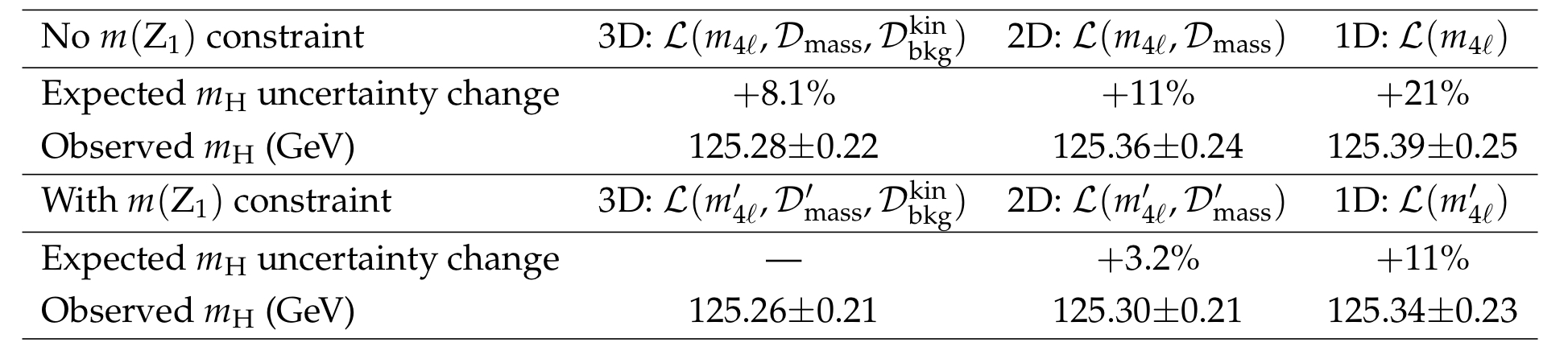}
\caption{A summary of the precision gained by refitting the lepton $\pt$ values which come from a mass constraint on the $\massof{Z}{1}$ (comparison across rows) and by introducing a 1D, 2D, or 3D fit likelihood fit (comparison across columns) using the observables described in Section~\ref{sec:observables}.
The greatest precision gain is observed when comparing a 1D likelihood fit (with no mass constraint) to a 3D fit (with a mass constraint) which results in a 21\% gain in precision on the Higgs boson mass measurement~\cite{HIG-16-041}. 
} 
\label{tab:table_improvement}
\end{table}
The Higgs boson mass was measured to be $\mh = \mhvallong$,
which achieved a better overall precision than the previous $\mh$ measurement of $\mhvalrunonelong$, made by the ATLAS and CMS Collaborations, which corresponded to an integrated luminosity of approximately $\fb{25}$ per collaboration from LHC proton-proton collision data in 2010 and 2011.
The 1D likelihood scan vs. $\mh$ accounting for one, two, and three observables is shown in Fig.~\ref{fig:likelihood_scan} (Left) and split up into different final states: $\fs$ (Right)~\cite{HIG-16-041}.
Finally, the precision gained by using 1D, 2D, and 3D likelihood fits, with and without a mass constraint on the $\massof{Z}{1}$, is given in Table~\ref{tab:table_improvement}~\cite{HIG-16-041}.
It should be noted that while these these proceedings were being written, an even more precise measurement of the Higgs boson mass was made public~\cite{HIG_19_004}. 

\section*{Acknowledgements}
This work is presented on behalf of the CMS collaboration. 
The author is grateful to Prof. Andrey Korytov, for the opportunity to present at DPF, to Dr. Hualin Mei, for having developed the bulk of the analysis, and to Dr. Filippo Errico for his limitless patience.

\end{document}